\documentclass[%
reprint,
nofootinbib,
 amsmath,amssymb,
 aps,
]{revtex4-2}

\usepackage{graphicx}% Include figure files
\usepackage{dcolumn}% Align table columns on decimal point
\usepackage{bm}% bold math
\usepackage{hyperref}% add hypertext capabilities
\usepackage[mathlines]{lineno}% Enable numbering of text and display math
\usepackage{amsmath,amssymb,euscript}
\usepackage{calrsfs}

\usepackage{tikz}
\usetikzlibrary{calc,decorations.markings}
\usetikzlibrary{snakes}
\usepackage{rotating}
\usepackage{comment}
\usepackage{empheq}

\newcommand{\eul}{\mathrm{e}}

\newcommand{\dd}{\mathrm{d}}

\newcommand{\Op}{\mathcal{O}}

\newcommand{\zb}{\bar{z}}
\newcommand{\Lie}[2]{\mathcal{L}_{#1}#2}

\allowdisplaybreaks[2]
\numberwithin{equation}{section}

\definecolor{darkgreen}{rgb}{0.0, 0.55, 0.1}

\begin{document}

%\preprint{LMU-ASC 03/21}

\title{Asymptotic dynamics and charges for FLRW spacetimes}% Force line breaks with \\

\author{Mart\'in Enr\'iquez Rojo}
\email{martin.enriquez@physik.lmu.de}    
 \affiliation{Arnold-Sommerfeld-Center for Theoretical Physics,\\ 
			Ludwig-Maximilians-Universit\"at of Munich,\\
			Theresienstr. 37, D-80333 M\"unchen, Germany}%Lines break automatically or can be forced with \\
\author{Till Heckelbacher}%
 \email{till.heckelbacher@physik.lmu.de}
\affiliation{Arnold-Sommerfeld-Center for Theoretical Physics,\\ 
			Ludwig-Maximilians-Universit\"at of Munich,\\
			Theresienstr. 37, D-80333 M\"unchen, Germany}%
\author{Roberto Oliveri}%
    \email{roliveri@fzu.cz}%
    \affiliation{CEICO, Institute of Physics of the Czech Academy of Sciences,\\ Na Slovance 2, 182 21 Praha 8, Czech Republic}

\begin{abstract}
We investigate the asymptotia of decelerating and spatially flat FLRW spacetimes at future null infinity. 
We find that the asymptotic algebra of diffeomorphisms can be enlarged to the recently discovered Weyl-BMS algebra for asymptotically flat spacetimes by relaxing the boundary conditions. This algebra remains undeformed in the cosmological setting contrary to previous extensions of the BMS algebra.
We then study the equations of motion for asymptotically FLRW spacetimes with finite fluxes and show that the dynamics is fully constrained by the energy-momentum tensor of the source. Finally, we propose an expression for the charges that are associated with the cosmological supertranslations and whose evolution equation features a novel contribution arising from the Hubble–Lema\^itre flow.
\end{abstract}

\maketitle
\newpage
%\tableofcontents

%==========================
	%====== INTRODUCTION  =====
	%==========================
	\section{Introduction}  
	\label{Sec:Intro}

	The study of the asymptotic region of an isolated self-gravitating source dates back to the pioneering work of Bondi, van der Burg, Metzner and Sachs (BMS)~\cite{Bondi:1962px,Sachs:1962wk,Sachs:1962zza}. These works initiated a rigorous research program to study gravitational waves in asymptotically flat spacetimes; see, \emph{e.g.}, \cite{Madler:2016xju,Strominger:2017zoo,Compere:2019qed} for a review.
	Over the past years, there has risen some interest in the asymptotia of a cosmological setting, with a special focus on decelerating and spatially flat Friedmann–Lema\^itre–Robertson–Walker
	(FLRW) spacetimes. These geometries are endowed with a future null infinity and, in addition, are employed to describe the radiation- and matter-dominated epochs in the evolution of the universe \cite{Mukhanov:2005sc}.
	The geometrical foundations of decelerating and spatially flat FLRW spacetimes at future null infinity have been initiated in  \cite{Bonga:2020fhx,Rojo:2020zlz,Enriquez-Rojo:2021blc}. 
	There are several reasons to perform and deepen into these studies. From a phenomenological point of view, it is essential for the transition from asymptotically flat toward cosmological spacetimes, and FLRW is the most natural candidate to begin with. It also proves rewarding to investigate whether the increasingly refined technical tools and relations, introduced in the context of asymptotically flat spacetimes, hold in more realistic scenarios. A prominent example is to discern whether the infrared triangle \cite{Strominger:2017zoo,Strominger:2014pwa} connecting asymptotic symmetries, soft theorems and memory effects in asymptotically flat spacetimes survives in cosmological spacetimes and, either way, which are the possible modifications and interpretation.
	Rather astoundingly, the literature regarding the infrared structure of cosmological spacetimes is very limited. The first attempt belongs to Hawking who proposed that the asymptotic symmetry group of asymptotically FLRW spacetimes reduces to its global symmetry group \cite{Hawking:1968qt}. Nevertheless, only very specific dust-filled universes with negative spatial curvature were considered, while the most recent studies \cite{Bonga:2020fhx,Rojo:2020zlz,Enriquez-Rojo:2021blc} treat spatially flat universes allowing for general matter content. 
	In the past years, several related studies have been performed in various directions: from the study of FLRW at timelike infinity \cite{Shiromizu:1999iq} to the asymptotic symmetries with non-vanishing cosmological constant \cite{Ferreira:2016hee,Compere:2019bua,Compere:2020lrt,Chrusciel:2021ttc}; and from the relation between adiabatic modes and soft theorems \cite{Hinterbichler:2013dpa,Mirbabayi:2016xvc,Pajer:2017hmb,Hamada:2018vrw} to 
	memory effects in de Sitter and $\Lambda$CDM cosmologies \cite{Bieri:2015jwa,Tolish:2016ggo,Chu:2016qxp,Chu:2016ngc,Bieri:2017vni,Hamada:2017gdg,Chu:2021apx}.

	In this work, we push forward the most recent studies \cite{Bonga:2020fhx,Rojo:2020zlz,Enriquez-Rojo:2021blc} on asymptotically decelerating and spatially flat FLRW spacetimes at future null infinity in two principal directions.

	On the one hand, from a purely geometrical perspective -- and motivated by the recent extension of the asymptotic algebra of diffeomorphisms in asymptotically flat spacetimes \cite{Freidel:2021fxf} denoted Weyl-BMS algebra 
	-- we relax the strong Bondi gauge and allow the diffeomorphisms to change the determinant of the metric on the celestial sphere.
	The asymptotic algebra turns out to be isomorphic to the Weyl-BMS algebra, in contrast to the one-parameter deformations of the BMS and generalized BMS algebras introduced in \cite{Enriquez-Rojo:2021blc}. This shows that the Weyl-BMS algebra is more rigid to deformations than the other extensions.

	On the other hand, we focus on the dynamics and develop the first on-shell analysis for these cosmological asymptotic metrics in General Relativity by investigating the asymptotic Einstein equations. 
	In particular, we explicitly solve the equations of motion for a subclass of metrics compatible with the supertranslation-like diffeomorphisms. The resultant analysis shows that the dynamics at future null infinity is completely determined in terms of the energy-momentum tensor, contrary to asymptotically flat spacetimes, where the Bondi news is unconstrained and the tensor degrees of freedom propagate. 

	Finally, as a third result and benefiting from the previous analysis, we propose suitable candidates for supertranslation-like charges in certain simplified settings whose  evolution  involves  a novel  Hubble  term  compared to asymptotically flat spacetimes. 

	The structure of this paper is as follows. In section \ref{Sec:ReviewFLRW}, we briefly review asymptotically FLRW spacetimes from the perspective of \cite{Rojo:2020zlz,Enriquez-Rojo:2021blc}. In section \ref{Sec:AsymptoticSymmetry}, we allow for Weyl transformations and obtain the asymptotic algebra of diffeomorphisms.
	Adopting General Relativity as our gravity theory, in section \ref{Sec:OnShellanalysis}, we develop an on-shell analysis of our cosmological spacetimes, with a special emphasis on the subset of metrics consistent with the absence of Weyl diffeomorphisms. This subset of metrics is used in section \ref{Sec:charges}, where we introduce charges for the supertranslation-like asymptotic diffeomorphisms. We conclude with a summary of results and future research in section \ref{Sec:Discussion}. Finally, we relegate the asymptotic Lie derivatives and a complementary analysis of the Weyl scalars for our metrics to the appendices \ref{Sec:AppLieDer} and \ref{Sec:Weylscalars}, respectively.

    \smallskip
	\textbf{Notation:} We generally use ``mathfrak'' font for the algebras, \emph{e.g.} $\mathfrak{bms}$ for the BMS algebra. Indices on the sphere are denoted by capital latin letters $A,B,C,...$. These indices are raised and lowered with the leading term $q_{AB}$ of the expansion of the metric on the sphere. $D_A$ denotes the covariant derivative with respect to $q_{AB}$. The Ricci scalar on the two-sphere is denoted by $\mathcal{R}$, while $R_{flat}$ and $R^{\text{FLRW}}$ denote the Ricci scalar on the four-manifold of asymptotically flat and exact FLRW spacetime. $\triangle G_{\mu\nu}\equiv G_{\mu\nu}-G_{\mu\nu}^{\text{FLRW}}$ stands for the difference between the Einstein tensor of asymptotically FLRW and exact FLRW. We use $\delta$ for the variations along the phase space, \emph{e.g.} $\delta_{f}$ denotes the action on the phase space of a vector field generated by $f$. The Hubble scale is given by $H=\partial_u a$, where $a$ is the conformal expansion scale factor of FLRW.

	%==========================
	%====== REVIEW of FLRW  =====
	%==========================
	
	\section{Review of asymptotically FLRW spacetimes}
	\label{Sec:ReviewFLRW}
	We briefly review the asymptotia of spatially flat FLRW and the treatment of asymptotically decelerating spatially flat FLRW universes at future null infinity $\mathcal{I}^+$. We refer the reader to \cite{Rojo:2020zlz,Enriquez-Rojo:2021blc} for more details.

	\subsection{FLRW spacetimes and their asymptotia}
	
	The metric of spatially flat FLRW spacetimes is given by
	\begin{align}
		\dd \bar{s}^2 &= - \dd t^2 + a^2(t)\left(\dd r^2+r^2\dd \Omega_{S^2}\right) \ ,\\* 
		a(t)&=\left(\frac{t}{t_0}\right)^{\frac{2}{3(w+1)}} \nonumber\ ,
	\end{align}
    and is sourced by a perfect fluid 
    \begin{equation}
        \bar{T}_{\mu\nu} = \left(e +p\right)\bar{u}_{\mu}\bar{u}_{\nu} + p \bar{g}_{\mu\nu}\,,\label{eq:EMT_background_cart}
    \end{equation}
    where $\bar{u}^{\mu} = \{1,0,0,0\}$ is the fluid four-velocity in the comoving frame, $e \propto a^{-3(w+1)}$ is the energy density, $p$ is the pressure and  they are related by the equation of state $p = w e$ with $w$ being a real constant.
    
	These metrics are related to the Minkowski metric by a Weyl transformation. Indeed, using the conformal time $\dd \eta= \dd t/a(t)$ and Bondi coordinates
	\begin{align}
		u&=\eta-\sqrt{x^ix_i} \ , \ \ r=\sqrt{x^ix_i} \ ,\nonumber\\ 
		z&=\frac{x^1+ix^2}{x^3+\sqrt{x^ix_i}} \ , \ \ \zb=\frac{x^1-ix^2}{x^3+\sqrt{x^ix_i}} \ ,
	\end{align}
	the spatially flat FLRW metric reads as
	\begin{align}
		\dd s^2 &=a^2(u,r)\left(-\dd u^2-2\dd u\dd r+\frac{4r^2}{(1+z\zb)^2}\dd z\dd\zb\right) \nonumber\\*
		&a(u,r)=\left(\frac{r+u}{L}\right)^{k}\label{pureFLRW} \ ,
	\end{align}
	where $L$ is a length scale and $k=2/(3w+1)$. 

\begin{figure*}
		\begin{center}
            \includegraphics[scale=0.3]{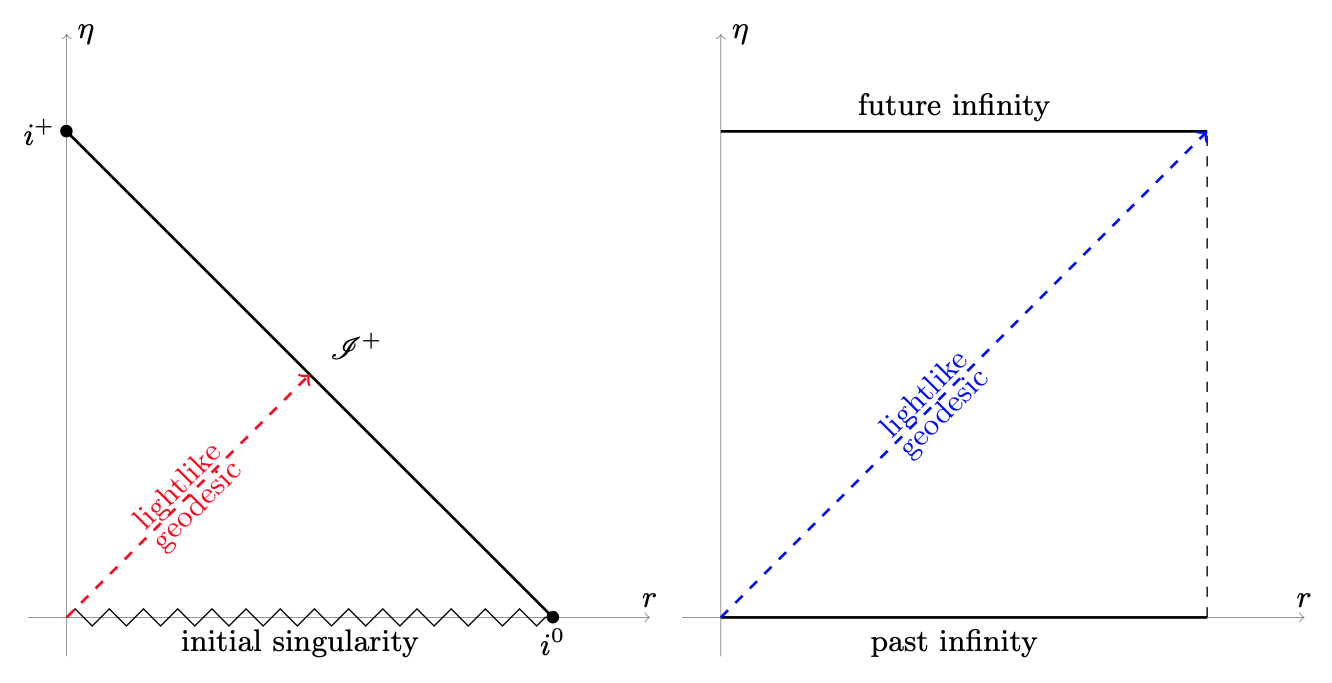}
		\end{center}
		\caption{Penrose diagram of spatially flat decelerating FLRW (left) and accelerating FLRW (right)}
		\label{fig:conformal_diagram}
	\end{figure*}

	These spacetimes can be divided into decelerating ($k>0$) and accelerating ($k<0$). The corresponding Penrose diagrams (see \emph{e.g.}, \cite{Mukhanov:2005sc,Harada:2018ikn}) are shown in figure \ref{fig:conformal_diagram}.
	
	Comparing the asymptotic regions of a light-like geodesic, it is clear that only decelerating FLRW spacetimes have a future null infinity $\mathcal{I}^+$. 
	For this reason, we will restrict ourselves to decelerating universes in this paper, leaving the investigation of accelerating FLRW spacetimes for future work.

	Finally, the non-vanishing components of the Einstein tensor of the exact FLRW background \eqref{pureFLRW} are given by
	\begin{align}
		G_{uu}^{\text{FLRW}}&=\frac{3k^2}{(u+r)^2}=\frac{3 k^2}{r^2}+\Op(r^{-3})\nonumber \ ,  \\
		G_{ur}^{\text{FLRW}}&=\frac{3k^2}{(u+r)^2}=\frac{3 k^2}{r^2}+\Op(r^{-3}) \ ,  \\
		G_{rr}^{\text{FLRW}}&=\frac{2k(1+k)}{(u+r)^2}=\frac{2 k (k+1)}{r^2}+\Op(r^{-3})\nonumber \ , \\
		G_{z\zb}^{\text{FLRW}}&=-\frac{r^2k(k-2)\gamma_{z\zb}}{(u+r)^2}=-\gamma_{z\zb}k(k-2)+\Op(r)\label{eq:Einstein_tensor_FLRW} \ . \nonumber 
	\end{align}

    The energy-momentum tensor in Bondi coordinates for a perfect fluid is easily obtained from \eqref{eq:EMT_background_cart} and is given by
    \begin{align}
        T_{\mu\nu}=a^2\begin{pmatrix}
            e & e & 0 & 0\\
            e & e+p & 0 & 0\\
            0 & 0 & 0& p r^2\gamma_{z\bar z} \\
            0 & 0& p r^2\gamma_{z\bar z} & 0
        \end{pmatrix}\,.
    \end{align}
    Since the energy density scales like $e\propto a^{-3(w+1)}$ and the evolution of the scale factor in terms of conformal time is given by $a\propto\eta^{\frac{2}{3w+1}}$, the energy-momentum tensor overall behaves as
    \begin{align}
        T_{\mu\nu}=\frac{a^2_0 e_0}{(u+r)^2}\begin{pmatrix}
            1 & 1 & 0 & 0\\
            1 & \frac{2(k+1)}{3k} & 0 &0\\
            0 & 0 &0 & \frac{2-k}{3k} r^2\gamma_{z\bar z}\\
            0 & 0 & \frac{2-k}{3k} r^2\gamma_{z \bar z}&0
        \end{pmatrix}\,,
    \end{align}
    which is consistent with the Einstein tensor \eqref{eq:Einstein_tensor_FLRW}.

	\subsection{Asymptotically decelerating and spatially flat FLRW spacetimes}
	
	In this section, we briefly recapitulate the ansatz and results for asymptotically decelerating spatially flat FLRW spacetimes obtained in our previous works \cite{Rojo:2020zlz,Enriquez-Rojo:2021blc}.
	
	\subsubsection{Working ansatz}
	
	To define which class of spacetimes asymptotes to decelerating and spatially flat FLRW at $\mathcal{I}^+$, the following conditions have been imposed in \cite{Rojo:2020zlz,Enriquez-Rojo:2021blc}:
	\begin{enumerate}
		\item The background metric, that is the metric in which all the asymptotic expansion coefficients vanish, is the exact FLRW in Eq.~\eqref{pureFLRW}.
		\item The strong Bondi gauge and frame are satisfied, meaning that
		\begin{align}
			\label{eq:Bondi_gauge}
			g_{rr}=0 \ , \quad  g_{rA}&=0 \ , \quad \partial_r\det\left(\frac{g_{AB}}{a^2 r^2}\right)=0 \ ,\\
			&\delta\sqrt{\det(g_{AB})}=0 \ ,\nonumber
		\end{align}
		where the indices $A,B\in\{z,\zb\}$ label the angular coordinates. These conditions will be preserved by the action of the asymptotic symmetries; see App.\ref{Sec:AppLieDer} for details. To be precise, the first three equations are gauge conditions, while the fourth one is a boundary condition on the celestial sphere. The latter can be relaxed as we shall see in section \ref{Sec:AsymptoticSymmetry}. 
		\item Allowance of cosmological perturbations preserves (to leading order) homogeneity, isotropy and spatial flatness, and leaves the equation of state of the background fluid invariant in the limit $r\to\infty$.
		\item The boundary conditions are preserved, meaning that no overleading terms are generated in the $r$ expansion upon application of asymptotic diffeomorphisms.
		\item Trace and components of the Einstein tensor cannot diverge in the limit $r\to\infty$, when integrated over the comoving sphere. Assuming General Relativity, these conditions translate directly into certain requirements for the energy-momentum tensor.
        In particular, we assume the following fall-off conditions of the energy-momentum tensor
        \begin{align}\label{falloffT}
            T_{uu} &= \mathcal{O}(r^{-2}),\,
            &T_{ur} &= \mathcal{O}(r^{-2}),\,
            &T_{uA} &= \mathcal{O}(r^{-1}),
            \nonumber\\
            T_{rr} &= \mathcal{O}(r^{-2}),\,
            &T_{rA} &= \mathcal{O}(r^{-1}),\nonumber\\
            T_{AB} &= \mathcal{O}(1)\,.
        \end{align}
	\end{enumerate}
	These considerations led to the following class of metrics \cite{Enriquez-Rojo:2021blc}~\footnote{Note that the sign of the coefficients in the $dudx^A$ part of the metric follows the convention of \cite{Rojo:2020zlz,Enriquez-Rojo:2021blc} and is the opposite to the sign convention in asymptotically flat spacetimes.}
	\begin{widetext}
	\begin{align}\label{eq:metric_allButPhi_uindependent}
		\dd s^2
		=\left(\frac{r+u}{L}\right)^{2k}\Bigg\{&-\left(1-\Phi-\frac{2m}{r}\right)\dd u^2-2\left(1-\frac{K}{r}\right)\dd u\dd r-2\left(r\Theta_A+U_A+\frac{1}{r}N_A\right)\dd u\dd x^A \nonumber  \\*
		&+\left(r^2q_{AB}+rC_{AB}+\mathcal{D}_{AB}+\frac12C_{AC}C^C_B\right)\dd x^A\dd x^B + \dots \Bigg\}.
	\end{align}
	\end{widetext}
	It represents an expansion in powers of $1/r$ for $r\to\infty$, where all the expansion coefficients are functions of $u,z$ and $\zb$, except for $q_{AB}$ which only depends on the angular coordinates $z$ and $\zb$. 
	Before continuing, let us point out that the ansatz \eqref{eq:metric_allButPhi_uindependent}, as well as the asymptotic diffeomorphisms preserving it, has been proven to give the correct flat limit when $k\to0$ in \cite{Enriquez-Rojo:2021blc}. Furthermore, $\Phi$, $m$ and $K$ transform as scalars under spatial rotations while $\Theta_A, U_A$ and $N_A$ transform as vectors, and $q_{AB}$, $C_{AB}$ and $\mathcal{D}_{AB}$ as tensors. The determinant condition in Eq.~\eqref{eq:Bondi_gauge} implies $C_{AB}$ and $\mathcal{D}_{AB}$ to be traceless. By comparing the expansion \eqref{eq:metric_allButPhi_uindependent} to the asymptotically flat expansion, we expect the parameter $m$ to be related to the mass of a central inhomogeneity, $C_{AB}$ to the gravitational radiation and $N_A$ to the angular momentum aspect of the spacetime. However, it is important to stress that the treatment so far has mostly been off-shell and that different coefficients do not yet have a sharp physical interpretation.
	In \cite{Enriquez-Rojo:2021blc} it is also shown that the ansatz \eqref{eq:metric_allButPhi_uindependent} naturally includes white holes but, to include simple cosmological black hole metrics like Sultana-Dyer, Thakurta and Vaidya, the expansion in $1/r$ has to be extended with logarithmic terms. As expected, the logarithmic ansatz does not generally satisfy the peeling property but preserves the asymptotic algebra.\footnote{We remark that the logarithmic terms enter at subleading order and, therefore, should be included in our on-shell analysis of section \ref{Sec:OnShellanalysis} and adequately treated. Such an analysis is beyond the scope of this paper, but we expect that it will not distort the essence of the results contained herein.}
	In addition, we comment that a $u$-dependent metric on the sphere $q_{AB}$ would imply $a^{-2}g_{uu}\propto \mathcal{O}(r)$ because of the closure of the metric under the action of the asymptotic diffeomorphisms. However, this term is not compatible with the third condition leading to our ansatz. 
	
	These observations play an important role in the forthcoming on-shell analysis of section \ref{Sec:OnShellanalysis}.

	\subsubsection{Asymptotic algebra of diffeomorphisms}
	
	The asymptotic diffeomorphisms and their action on the asymptotic data for the class of metrics in Eq.~\eqref{eq:metric_allButPhi_uindependent} have been computed in the case of local conformal Killing vectors (CKV)-superrotations \cite{Rojo:2020zlz} and in the case of $\text{Diff}(S^2)$ diffeomorphisms \cite{Enriquez-Rojo:2021blc}. In section \ref{Sec:AsymptoticSymmetry}, we will allow for local Weyl transformations.
	For the time being, it is instructive to review the structure of the asymptotic algebra at future null infinity $\mathcal{I}^+$. In such a limit, and using the new parameter $(1+s)\equiv(1+2k)/(1+k)$, the asymptotic diffeomorphisms become
	\begin{multline}
		\xi[f(z,\zb),V^A(z,\zb)]=\\
		=\left(f+\frac{u}{2}(1+s)D_AV^A\right)\partial_u+V^A\partial_A \ ,
		\label{eq:universal_diffeo_1}
	\end{multline}
	where $f(z,\zb)$ and $V^A(z,\zb)$ denote, respectively, supertranslation- and superrotation-like transformations.
	Their Lie bracket gives
	\begin{equation}
		\xi[\hat{f},\hat{V}^A]=\Big[\xi[f,V^A],\xi[f',V'^A]\Big],
	\end{equation}
	where the hatted gauge parameters read as
	\begin{align}
		\hat{f}=&V^AD_Af'-V'^AD_Af \nonumber\\
		&+\frac{(1+s)}{2} \left(fD_AV'^A -f'D_AV^A\right) \ , \\
		\hat{V}^A=&V^BD_BV'^A-V'^BD_BV^A \ .
		\label{eq:algebra_1}
	\end{align}
	We obtain a one-parameter deformation of the extended BMS algebra \cite{Barnich:2010eb,Barnich:2016lyg} denoted as $\mathfrak{b}\mathfrak{m}\mathfrak{s}_s\simeq(\mathfrak{w}\mathfrak{i}\mathfrak{t}\mathfrak{t}\oplus\mathfrak{w}\mathfrak{i}\mathfrak{t}\mathfrak{t})\ltimes_s\mathfrak{s}_s$, where the vectors $V^A$ are local CKV on $S^2$, and a deformation of the generalized BMS algebra \cite{Campiglia:2015yka,Campiglia:2014yka} denoted as $\mathfrak{g}\mathfrak{b}\mathfrak{m}\mathfrak{s}_s\simeq\mathfrak{vect}(S^2)\ltimes_s\mathfrak{s}_s$, where the vectors $V^A$ are smooth diffeomorphisms on the sphere.~\footnote{$\mathfrak{vect}(S^2)$ denotes the algebra of globally defined vector fields on the sphere.} Both reduce to a one-parameter deformation of the original BMS algebra $\mathfrak{b}_s\simeq\mathfrak{s}\mathfrak{o}(1,3)\ltimes\mathfrak{s}_s$, found in \cite{Bonga:2020fhx}, when restricting to the six $V^{A}$ that are global CKV on $S^2$.~\footnote{For a comparison between our results and those of \cite{Bonga:2020fhx}, we refer the reader to \cite{Rojo:2020zlz,Enriquez-Rojo:2021blc}.}
	
	These algebras are clearly one-parameter deformations of the original, extended and generalized BMS algebras, where the deformation parameter $s$ is directly related to the equation of state of the background fluid and unveils a cosmological holographic flow deformation at the level of the asymptotic algebras. In section \ref{subsec_defwbms}, we will notice that the deformation of the Weyl-BMS algebra becomes trivial when we allow for Weyl transformations.
	
	As a final comment, we briefly note that in \cite{Enriquez-Rojo:2021blc} it was pointed out that the deformed extended BMS algebra $\mathfrak{bms}_s$ corresponds to the element $W\left(-\frac{1+s}{2},-\frac{1+s}{2};-\frac{1+s}{2},-\frac{1+s}{2}\right)$ of the four-parametric family of deformations of $\mathfrak{bms}$, denoted by $W(a,b;\bar{a},\bar{b})$ \cite{Safari:2019zmc,Safari:2020pje}. Furthermore, it was shown in \cite{Enriquez-Rojo:2021rtv} that, after a change of topology from $S^2$ to the doubly punctured plane $\mathbb{C}^{*}$, the deformed generalized BMS algebra $\mathfrak{gbms}_s$ can be viewed as the member $gW\left(-\frac{1+s}{2},-\frac{1+s}{2};-\frac{1+s}{2}\right)$ of the three-parametric family of deformations of $\mathfrak{gbms}$, called $gW\left(a,b;\bar{a}\right)$ \cite{Enriquez-Rojo:2021rtv}. 
	
	%================================
	%=== ASYMPTOTIC SYMMETRIES  =====
	%================================
	\section{BMSW-like diffeomorphisms in FLRW}
	\label{Sec:AsymptoticSymmetry}
	
	In this section, we allow for Weyl-BMS transformations in asymptotically decelerating and spatially flat FLRW spacetimes, following the corresponding treatment in asymptotically flat spacetimes \cite{Freidel:2021fxf}. 
	In the rest of the paper, we will assume that the leading asymptotic coefficients $\Phi$, $\Theta_A$ and $q_{AB}$ are $u$-independent. This choice implies finite fluxes through the boundary and will be motivated by our on-shell treatment in section \ref{Sec:OnShellanalysis_finitefluxes}.

	\subsection{Residual transformation in Bondi gauge}
	
	We analyze the residual diffeomorphisms for the on-shell metrics \eqref{eq:metric_allButPhi_uindependent}  starting from 
	\begin{multline}
		\xi=\xi^u(u,z,\zb)\partial_u+\left[r\xi^{r(V)}(z,\zb)+\xi^{r(0)}+\frac{1}{r}\xi^{r(1)} +\dots\right]\partial_r\\
		+\left[V^{B}(z,\zb)+\frac1r\xi^{B(1)}+\frac{1}{r^2}\xi^{B(2)} +\dots\right]\partial_B \ ,
		\label{eq:Srotacc}
	\end{multline}
    where dots stand for subleading terms in $1/r$ that enter the $\Op(r^{-2})$ in $a^{-2}\mathcal{L}_{\xi}g_{\mu\nu}$ in App.\ref{Sec:AppLieDer}.
	We emphasize that, contrary to previous works \cite{Rojo:2020zlz,Enriquez-Rojo:2021blc}, we do not require the determinant of the metric on the sphere to be fixed. Instead of the strong Bondi gauge, we follow \cite{Freidel:2021fxf} and use the Bondi gauge
	\begin{equation}
		g_{rr}=0,\quad g_{rA}=0,\quad \partial_r\det\left(\frac{g_{AB}}{a^2 r^2}\right)=0 \ .
	\end{equation}

	The condition on $\Lie{\xi}{g_{rr}}$ is already verified by the ansatz. The vanishing of $\Lie{\xi}{g_{rA}}$ leads to the following restrictions:
	\begin{align}
		\xi_A^{(1)}&=-D_A\xi^u \ , \\
		\xi_A^{(2)}&=\frac12\left(KD_A\xi^u-C_{AB}\xi^{B(1)}\right) \ .
	\end{align}
	To satisfy the determinant condition, we have to demand that $q^{AB}C_{AB}=0$, $q^{AB}S_{AB}=C^{AB}F_{AB}$ and that $q^{AB}K_{AB}=C^{AB}S_{AB}-C^A_CC^{CB}F_{AB}+(\mathcal{D}^{AB}+\frac12C^A_C C^{CB})F_{AB}$, where $K_{AB}$, $S_{AB}$ and $F_{AB}$ are defined in \eqref{eq:LieAB11srot}. This leaves the leading order contribution to the spherical metric arbitrary, which means that the coefficient $\xi^{r(V)}$ in the expansion \eqref{eq:Srotacc} joins $f$ and $V^A$ as a free parameter. Besides, we obtain
	\begin{widetext}
	\begin{align}
		\xi^{r(0)}&=\frac{1}{1+k}\left[-\frac{1}{2}D_A\xi^{A(1)}-\frac{1}{2}\Theta^AD_A\xi^u+ku\xi^{r(V)}-k\xi^u\right] \ ,
		\\
		\xi^{r(1)}&=\frac{1}{2(1+k)}\left[C^A_B\Theta_AD^B\xi^u-2k\left(u^2\xi^{r(V)}-u\xi^{r(0)}-u\xi^u\right)-D_A\xi^{A(2)}+U^AD_A\xi^u\right] \ .
	\end{align}
	\end{widetext}
	The remaining requirements come from $\Lie{\xi}{g_{uA}}=\Op(r)$, $\Lie{\xi}{g_{uu}}=\Op(1)$ and $\Lie{\xi}{g_{ur}}=\Op(r^{-1})$. Altogether they translate into
	\begin{align}
		\partial_uV^A&=\partial_u\xi^{r(V)}=0\\
		\partial_u\xi^u&=-(1+2k)\xi^{r(V)}\\
		\Rightarrow&\quad\xi^u=f(z,\zb)-u(1+2k)\xi^{r(V)}(z,\zb) \ .
	\end{align}
	
	\subsection{Asymptotic algebra} %- Deformation of Weyl-BMS}
	\label{subsec_defwbms}
	
	At $r\to\infty$, $r=\text{constant}$,  our diffeomorphisms become
	\begin{multline}
		\xi[f(z,\zb),\xi^{r(V)}(z,\zb),V^A(z,\zb)]\\
		=\left[f-u(1+2k)\xi^{r(V)}\right]\partial_u+V^A\partial_A \ ,
		\label{eq:universaldiffeoxibmswk}
	\end{multline}
	leading to the asymptotic algebra
	\begin{align}
		V_{12}:=&[V_1, V_2]_{\text{Lie}} \ , \\
		\xi^{r(V)}_{12}=&V_1[\xi^{r(V)}_2]-V_2[\xi^{r(V)}_1] \ ,  \\  
		f_{12}=&V_1[f_2]-V_2[f_1]\nonumber\\
		&-(1+2k)(f_1\xi^{r(V)}_2-f_2\xi^{r(V)}_1) \ . 
	\end{align}
	
	We see that $f$ and $\xi^{r(V)}$ transform as scalars under $\text{Diff}(S^2)$, while $f$ also transforms as a weight-$(1+2k)$ section of the scale bundle. An alternative way to visualize the algebra is to compute
	\begin{multline}
		\xi[\hat{f},\hat{\xi}^{r(V)},\hat{V}^A]\\
		=\left[\xi[f,\xi^{r(V)},V^A],\xi[f',\xi^{r(V)'},V'^A]\right],
	\end{multline}
	where
	\begin{align}
		\hat{f}=&V^AD_Af'-V'^AD_Af\nonumber\\
		&-(1+2k)[f\xi^{r(V)'}-f'\xi^{r(V)}] \ ,\nonumber \\
		\hat{V}^A=&V^BD_BV'^A-V'^BD_BV^A \ ,\label{eq:algebra1}\\
		\hat{\xi}^{r(V)}=& V^AD_A\xi^{r(V)'}-V'^AD_A\xi^{r(V)} \ .\nonumber
	\end{align}
	Thus, we obtained the algebra $\mathfrak{bmsw}_k\simeq(\mathfrak{vect}(S^2)\ltimes\mathfrak{w})\ltimes_k\mathfrak{s}$, which one would naively regard as a deformation of $\mathfrak{bmsw}$ obtained in \cite{Freidel:2021fxf}. Nevertheless, the fact that the Weyl generators $\xi^{r(V)}$ are independent of $V^A$ allows us to rescale the former such that the algebra $\mathfrak{bmsw}_k$ is isomorphic to the Weyl-BMS algebra $\mathfrak{bmsw}$. This differs from the $\mathfrak{bms}_s$ and  $\mathfrak{gbms}_s$ algebras, where the one-parameter deformation is non-trivial and cannot be removed by a simple rescaling of the generators. As a consequence, we observe that the $\mathfrak{bmsw}$ algebra is more universal because it is more rigid toward deformations than $\mathfrak{bms}$ and $\mathfrak{gbms}$. 
	
	Let us explore the algebra \eqref{eq:algebra1} in a different basis by embedding $\mathfrak{v}\mathfrak{e}\mathfrak{c}\mathfrak{t}(S^2)$ into $\mathfrak{v}\mathfrak{e}\mathfrak{c}\mathfrak{t}(\mathbb{C}^\star)$, changing the topology to admit two punctures at the poles. In this case, the vector field in (\ref{eq:universaldiffeoxibmswk}) can be expressed as
	\begin{align}
		\xi(f_{pq},0,0)&:=T_{p,q}=z^p\bar{z}^q\partial_u  \ ,  \\
		\xi(0,\xi^{r(V)}_{pq},0)&:=W_{p,q}=-(1+2k)  z^p\bar{z}^q~u\partial_u \ , \\
		\xi(0,0,V^z_{mn})&:=\mathcal{L}_{m,n}=-z^{m+1}\bar{z}^n\partial_z \ ,  \\
		\xi(0,0,V^{\bar{z}}_{mn})&:=\hat{\mathcal{L}}_{m,n}=-z^{m}\bar{z}^{n+1}\partial_{\bar{z}} \ .
	\end{align}
	In terms of this basis, we obtain the following non-vanishing commutators
	\begin{subequations}\label{bmswalgebra}
		\begin{align}
			[\mathcal{L}_{m,n},\mathcal{L}_{r,s}]&=(m-r)\mathcal{L}_{m+r,n+s} \ , \\ 
			[\hat{\mathcal{L}}_{m,n},\hat{\mathcal{L}}_{r,s}]&=(n-s)\hat{\mathcal{L}}_{m+r,n+s} \ , \\ 
			[\mathcal{L}_{m,n},\hat{\mathcal{L}}_{r,s}]&=-r\hat{\mathcal{L}}_{m+r,n+s}+n\mathcal{L}_{m+r,n+s} \ ,  \\
			[\mathcal{L}_{m,n},W_{p,q}]&=-pW_{p+m,q+n}  \ , \\
			[\hat{\mathcal{L}}_{m,n},W_{p,q}]&=-qW_{p+m,q+n}  \ ,   \\
			[\mathcal{L}_{m,n},T_{p,q}]&=-pT_{p+m,q+n}  \ , \\
			[\hat{\mathcal{L}}_{m,n},T_{p,q}]&=-qT_{p+m,q+n}  \ ,   \\
			[W_{m,n},T_{p,q}]&=(1+2k)T_{p+m,q+n}\label{eq:oneplustwok} \ . 
		\end{align}
	\end{subequations}
	It is now evident that the factor $(1+2k)$ in the last commutator can easily be removed by a rescaling of $W_{m,n}$, leading to the isomorphism $\mathfrak{bmsw}_k\simeq\mathfrak{bmsw}$.~\footnote{It would be very interesting to explore the family of linear deformations of $\mathfrak{b}\mathfrak{m}\mathfrak{s}\mathfrak{w}$, similar to $W(a,b;\bar{a},\bar{b})$ for $\mathfrak{bms}$ \cite{Safari:2019zmc,Safari:2020pje} and $gW(a,b;\bar{a},\bar{b})$ for $\mathfrak{gbms}$ \cite{Enriquez-Rojo:2021rtv}.}
	
	As a final comment, let us note that a very similar algebra to \eqref{bmswalgebra} with $\mathfrak{witt}$-superrotations instead of $\mathfrak{vect}(\mathbb{C}^{\star})$ has been uncovered in Eq.~(2.31) of \cite{Donnay:2016ejv}. There, the authors performed a near-horizon analysis where the surface gravity $\kappa$ plays exactly the same role as the factor $(1+2k)$ in Eq.~\eqref{eq:oneplustwok}. A major difference is that in their case $\kappa$ cannot be reabsorbed due to the fact that the value $\kappa=0$ is included, whereas in our case $(1+2k)\neq0$.
	We also note that, our parameter $k$ can be identified\footnote{We thank the anonymous referee for suggesting this relationship.} with the level of the conformal Carroll algebra \cite{Duval:2014lpa}. In particular, by explicit comparison of our BMSW-like vector field \eqref{eq:universaldiffeoxibmswk} and the conformal Carroll vector field of level $k^{C}$ (see Eq.(IV.8) in \cite{Duval:2014lpa} with $d=2$), we get $\nabla_A V^A k^{C} = -(1+2k)\xi^{r(V)}$.
	
	\subsection{Action of the asymptotic diffeomorphisms}
	\label{Sec:OnShell}
	
	For completion and posterior use, we give the explicit variations of the asymptotic coefficients under the asymptotic diffeomorphisms \eqref{eq:Srotacc}:
	\begin{widetext}
	\begin{align}
		\delta\Phi &=V^AD_A\Phi-2\partial_u\xi^{r(0)}-2k(1-\Phi)\xi^{r(V)}
		-2(1-\Phi)\partial_u\xi^u+2\Theta_A\partial_u\xi^{A(1)},\label{eq:dNgenrot}\\
		\delta m &=\xi^u\partial_u m+V^AD_Am-k(1-\Phi)\xi^u-\left[(1-2k)m-ku(1-\Phi)\right]\xi^{r(V)}-k(1-\Phi)\xi^{r(0)}\nonumber\\
		&\quad+K\partial_u\xi^{r(0)}-\partial_u\xi^{r(1)}+m\partial_u\xi^u+U_A\partial_u\xi^{A(1)}+\frac12\xi^{A(1)}D_A\Phi+\Theta_A\partial_u\xi^{A(2)},\label{eq:dmgenrot}\\
		\delta K &=\xi^u\partial_u K+V^AD_AK+K\partial_u\xi^u-\Theta_A\xi^{A(1)}+2k\left(u\xi^{r(V)}-\xi^u-\xi^{r(0)}\right)+2kK\xi^{r(V)},\label{eq:dFgenrot}\\
		\delta q_{AB}&=2(1+k)\xi^{r(V)}q_{AB}+\mathcal{L}_{V}q_{AB},\label{eq:dqABgenrot}\\
		\delta C_{AB} &=\xi^u\partial_uC_{AB}+\Lie{V}{C_{AB}}+(1+2k)C_{AB}\xi^{r(V)}
		+\Lie{\xi^{A(1)}}{q_{AB}}+\Theta_AD_B\xi^u+\Theta_BD_A\xi^u \nonumber\\
		&\quad+2q_{AB}\left[(1+k)\xi^{r(0)}-ku\xi^{r(V)}+k\xi^u\right],\label{eq:dCABgenrot}\\
		\delta \Theta_A&=\Lie{V}{\Theta_A}+(1+2k)\Theta_A\xi^{r(V)}
		-\partial_A\xi^{r(V)}+\Theta_A\partial_u\xi^u+q_{AB}\partial_u\xi^{B(1)},\label{eq:dAAgenrot}\\
		\delta U_A&=\xi^u\partial_u U_A+\Lie{V}{U_A}+\Lie{\xi^{C(1)}}{\Theta_A}+2k\Theta_A(\xi^u+\xi^{r(0)}-u\xi^{r(V)})-D_A\xi^{r(0)}+KD_A\xi^{r(V)}\nonumber\\
		&\quad -(1-\Phi)D_A\xi^u+U_A\partial_u\xi^u+C_{AB}\partial_u\xi^{B(1)}+2kU_A\xi^{r(V)}+\Theta_A\xi^{r(0)}+q_{AB}\partial_u\xi_B^{(2)},\label{eq:dUAgenrot}\\
		\delta N_A&=\xi^u\partial_u N_A+\Lie{V}{N_A}-(1-2k)N_A\xi^{r(V)}+N_A\partial_u\xi^u
		+\Lie{\xi^{C(1)}}{U_A}+\Lie{\xi^{C(2)}}{\Theta_A}\nonumber\\
		&\quad+KD_A\xi^{r(0)}-D_A\xi^{r(1)}+2mD_A\xi^u+2kU_A\left(\xi^{r(0)}+\xi^u-u\xi^{r(V)}\right)\nonumber\\
		&\quad+2k\Theta_A\left[u^2\xi^{r(V)}-u(\xi^{r(0)}+\xi^u)+\xi^{r(1)}\right]+\Theta_A\xi^{r(1)}+\left(\mathcal{D}_{AB}+\frac12C_{AC} C^C_B\right)\partial_u\xi^{B(1)} \nonumber \\
		&\quad+C_{AB}\partial_u\xi^{B(2)}\label{eq:dNAgenrot} \ .
	\end{align}
	\end{widetext}

	\section{Equations of motion}
	\label{Sec:OnShellanalysis}
	
	So far, we have reviewed the geometrical analysis performed in \cite{Rojo:2020zlz,Enriquez-Rojo:2021blc} and extended it in order to allow for Weyl transformations. Nevertheless, this treatment is off-shell, in the sense that we did not make explicit use of the equations of motion. In this section, we adopt General Relativity as our gravity theory and perform an on-shell analysis. 
	This means that we analyze the Einstein tensor as an expansion in $r^{-1}$, such that the expansion coefficients $G^{(i)}_{\mu\nu}$ are defined by
	\begin{align}
		G_{\mu\nu}=R_{\mu\nu}-\frac12 g_{\mu\nu}R=\sum\limits_{i} \frac{G_{\mu\nu}^{(i)}}{r^i}
	\end{align}
	and the Ricci scalar is expanded as 
	\begin{equation}
		R=\left(\frac{r+u}{L}\right)^{-2k}\sum\limits_{i=0}^\infty\frac{R^{(i)}}{r^i} \ .
	\end{equation}

    In the following, we compute the Einstein tensor and impose the fall-off behavior of the asymptotic FLRW energy-momentum tensor \eqref{falloffT} to find conditions on the metric functions and thus on the space of solutions.

	\subsection{Metrics with finite fluxes}
	\label{Sec:OnShellanalysis_finitefluxes}
	
	We begin by introducing the leading $uu$ and $uA$ components of the Einstein tensor obtained from the ansatz \eqref{eq:metric_allButPhi_uindependent}:
	\begin{align}
			G_{uu}^{(1)}=& -(1+k)\partial_u\Phi - q_{AB}(D^B\partial_u\Theta^A\nonumber\\
			&+(1+2k)\Theta^A\partial_u\Theta^B) \ ,\\
			G_{uA}^{(0)}=& -\frac{1}{2}\partial_u\Theta_A \ .
	\end{align}
	It can easily be observed that these components lead to linearly divergent fluxes at large $r$.\footnote{The presence of $u$-dependent leading terms, such as $\Phi(u)$, $\Theta_A(u)$ and $q_{AB}(u)$, would be necessary if one wants to describe dynamical perturbations of the FLRW boundary among our boundary metrics.}
	
	As a consequence, we restrict ourselves to the solutions in which these components vanish, which is equivalent to imposing $\partial_u\Phi=\partial_u\Theta_A=0$. This choice is consistent because the variations $\delta\Phi$ and $\delta\Theta_A$ generated by means of asymptotic transformations are $u$-independent if we start with $\Phi$ and $\Theta_A$ which do not depend on $u$, as can be quickly noticed from \eqref{eq:uuDiff} and \eqref{eq:uADiff}.
	
	The resulting metrics satisfy a series of properties that make them suited for a Bondi analysis.
	First, it is easy to notice that all the leading terms are $u$-independent, such that only the subleading terms can be dynamical. This is equivalent to taking as a boundary the equivalence class of unperturbed FLRW metrics allowed by $\mathfrak{bmsw}$ transformations, while the potential dynamics is restricted to the subleading terms $m$, $K$, $U_A$ and $C_{AB}$. The latter transform, respectively, as scalars, vector and tensor encoding (up to combinations) a maximum of six degrees of freedom, which can be reduced after imposing the remaining equations of motion.   
	Second, one can check that the resulting $G_{\mu\nu}$ components are of the same order in $r$ as the perfect fluid background, which is a reminiscence of the analysis performed in \cite{Bonga:2020fhx}. This guarantees that not only the $G_{\mu\nu}$ components but also their fluxes through future null infinity $\mathcal{I}^+$ are finite.

	\subsection{Asymptotic Einstein equations and degrees of freedom}
	
	Following the analysis of the previous subsection, we analyze the equations of motion and corresponding degrees of freedom for the on-shell ansatz \eqref{eq:metric_allButPhi_uindependent} with $\partial_u\Phi=\partial_u\Theta_A=\partial_u q_{AB}=0$. 
	
	\subsubsection{General case}
	In this section, we will present the leading Einstein equations and classify them in scalar, vector and tensor equations.
	
	\paragraph{Scalar equations}
	\label{Sec:Massloss}
	
	We start with the leading expression of $G_{uu}$
	\begin{widetext}
	\begin{align}\label{eq:Guu}
		G_{uu}^{(2)}&= \frac{1}{4}\partial_u C_{AB}\partial_u C^{AB} +D_A \partial_u U^A + 2(1+k)\partial_u \left( m+\Phi K\right)
		+\frac{1}{2}(q_{AC}q_{BD}-q_{AD}q_{BC})D^B\Theta^A D^D\Theta^C\cr 
		&\quad+ \partial_u K \left(2+2k\Theta^A \Theta_A + D_A \Theta^A \right)
		+\frac{1}{2}(1-2k)\Theta^A D_A \Phi -\frac{1}{2}\Delta \Phi + \Theta^A\left(D_A \partial_u K + 2k \partial_u U_A\right)\cr
		&\quad+(\Phi-1)\left[-\frac{1}{2}\mathcal{R}+\frac{1}{4}(1+8k+4k^2)\Theta_A\Theta^A+2(k+1)D_A\Theta^A\right]\cr
		&\quad-(\Phi-1)\left[(2k+1)(\Phi-1)+k^2(\Phi+1)\right] + 2k(k+1).
	\end{align}
	\end{widetext}
	This equation corresponds to the Bondi mass-loss equation in the asymptotically flat limit.
	
	The constraint equation for the parameter $K$ reads as
	\begin{equation}
		\begin{split}\label{eq:Grr3}
			& G_{rr}^{(3)}= -2(1+k)(2ku-K) \ .
		\end{split}
	\end{equation}
	Note that $K$ is completely fixed by the corresponding term in the expansion of the energy-momentum tensor.
	
	Besides, we also have 
	\begin{align}
		G_{ur}^{(2)}=&\frac12(\mathcal R-2)+3k^2+(1+k)^2\Phi\nonumber\\*
		&-\frac14(1+2k)^2\Theta_A\Theta^A+\frac12(3+4k)D_A\Theta^A \ ,
	\end{align}
	which does not generally impose any extra condition on the parameters.
	
	\paragraph{Vector equations}
	\label{Sec:vector}
	
	At leading order a novel constraint for the parameter $\Theta_A$ appears. It is given by
	\begin{equation}
		\begin{split}\label{eq:GrA1}
			& G_{rA}^{(1)}= (1+k)\Theta_A \ .
		\end{split}
	\end{equation}
	The function $\Theta_A$, just as $K$, is now completely determined by the corresponding expansion coefficient of the energy-momentum tensor.\\
	At subleading orders, we obtain the generalized version of the well-known constraint for $U_A$ in flat spacetimes
	\begin{align}
		G_{rA}^{(2)}=&\frac12\left[2ku\Theta_A-(3+2k)D_AK-D_BC_A^B\right.\nonumber\\*
		&\left.+(1+2k)\left(C_{AB}\Theta^B-2U_A - K\Theta_A\right)\right]
	\end{align}
	and
	\begin{widetext}
	\begin{align}
		G_{uA}^{(1)}&=\Theta_A \left[\frac12\mathcal R -1+\Phi-k(2-\Phi)+k^2(1+\Phi)\right]
		+\frac{\Theta_A}{4}\left[-(1+2k)^2\Theta_B\Theta^B+2(3+4k)D_B\Theta^B+6\partial_uK\right]\nonumber\\
		&\quad+\frac12\Big[-2kD_A\Phi+\partial_uD^BC_{AB}-D_BD_A\Theta^B+D_A D^A\Theta_A
		-\Theta^B(-2k(D_A\Theta_B-D_B\Theta_A)+(1+2k)\partial_uC_{AB})\nonumber\\
		&\qquad\quad+2\partial_uU_A+\partial_uD_AK\Big] \ ,
	\end{align}
	\end{widetext}
	which do not generally impose any new condition on the parameters.
	
	\paragraph{Tensor equations}
	\label{Sec:tensor}
	
	The leading order tensor components are given by 
	\begin{align}
		G_{AB}^{(0)}=&-\frac12\left[\Theta_A\Theta_B-(1+2k)(D_A\Theta_B+D_B\Theta_A)\right.\nonumber\\*
		&\left.-2k\partial_uC_{AB}\right]
		+\frac14q_{AB}\left[(4k^2-1)\Theta_C\Theta^C-4\partial_u K\right.\nonumber\\*
		&\left.-4(k(k(1+\Phi)-2)+(1+2k)D_C\Theta^C)\right],
	\end{align}
	which constitutes a novel constraint for the time evolution of $C_{AB}$ that is absent in asymptotically flat spacetimes. Interestingly, this condition, only present for $k\neq0$, is associated with the presence of a Hubble scale in expanding universes from which all the modes stop being oscillating and are frozen \cite{Mukhanov:2005sc}.
	
	\paragraph{Ricci scalar} Finally, let us analyze the value of the leading order Ricci scalar for our 
	\begin{align}\label{eq:Ricci}
		R^{(2)}=&\mathcal{R}-\frac32(1+2k)^2\Theta_A\Theta^A+2\left(-(1+3k)(1-\Phi)\right.\nonumber\\*
		&\left.+3k^2(1+\Phi)+
		(2+3k)D_A\Theta^A+\partial_uK\right) \nonumber \\
		=&R^{(2)}_{\text{FLRW}}+\left[\mathcal{R}-2-\frac32(1+2k)^2\Theta_A\Theta^A+2\partial_uK\right.\nonumber\\
		&\left.+2\left((1+3k+3k^2)\Phi+(2+3k)D_A\Theta^A\right)\right]
	\end{align}
	We would like to recall that in the flat limit, \emph{i.e.} $k\to0$, $\Theta_A\to0$ and $K\to0$, this equation becomes
	\begin{align}
		R_{flat}^{(2)}= 2\Phi+\mathcal{R}-2 \ .
	\end{align}
	In fact, the condition $R_{flat}^{(2)}\overset{!}{=}0$ is imposed as a flatness condition, leading to a constraint on $\Phi$; see \cite{Freidel:2021fxf}. Following the same logic, we can impose $R^{(2)}\overset{!}{=}R^{(2)}_{\text{FLRW}}$, which again constrains $\Phi$ in terms of $\mathcal{R}$, $T^{(1)}_{rA}$ and $T^{(3)}_{rr}$, determining a balance equation which ensures that the spacetimes under analysis still have an FLRW profile. 

	Before continuing, it is instructive to have a closer look at the values of the variations \eqref{eq:dNgenrot}, \eqref{eq:dFgenrot} and \eqref{eq:dAAgenrot} in our setting. In fact, we observe that they can be expressed as:
	\begin{align}
		\delta \Theta_A  = & \mathcal{L}_V \Theta_A +2 k \partial_A\xi^{r(V)} \ , \nonumber\\
		\delta \Phi  = & V^A D_A \Phi + \left[2(1-\Phi)(1+k)-4k\right.\nonumber\\*
		&\left.+\frac{1+2k}{1+k}\left(D_A D^A + (1+2k)\Theta^A D_A \right) \right] \xi^{r(V)} \ , \nonumber \\
		\delta K = & \xi^u\partial_uK+V^AD_AK-K\xi^{r(V)}+\frac{(1+3k)}{(1+k)}\Theta^AD_A\xi^u\nonumber\\
		&+\frac{2k}{(1+k)}[u\xi^{r(V)}-D_AD^A\xi^u-\xi^u] \ . \nonumber
	\end{align}
	These Lie derivatives confirm explicitly our previous statement that the choice $\partial_u\Phi=\partial_u\Theta_A=0$ is consistent because the variations $\delta\Phi$ and $\delta\Theta_A$ generated by means of asymptotic transformations are $u$-independent if we start with $\Phi$ and $\Theta_A$ not depending on $u$. Moreover, we observe that $\Theta_A$ is unavoidably generated by Weyl transformations, while in the presence of only supertranslations this component is not necessarily present. The same statement holds true for $\Phi$, whereas $K$ is generated in any case. Remarkably, in the absence of Weyl transformations, $K$ does not need to be $u$-dependent.
	
	\subsubsection{Absence of Weyl transformations}
	\label{subsec:onshellnoWeyl}
	
	In the last subsection, we have noticed how complicated the analytical treatment becomes in general settings. Nevertheless, the physical picture and the role of the different coefficients, as well as the nature of the different degrees of freedom, are exactly the same as in simpler backgrounds.~\footnote{Note that the backgrounds are encoded in the coefficients $\Theta_A$ and $\Phi$.} Therefore, we will now restrict ourselves to a simple setting, which is consistent with supertranslations and the absence of Weyl diffeomorphisms (\emph{i.e.} $\xi^{r(V)}=0$), with $\Theta_A=\Phi=0$ and analyze it in more detail, solving the Einstein equations explicitly. 
	
	Let us start by writing down the relevant Einstein equations \eqref{eq:Guu}-\eqref{eq:Ricci} in our simplified setting:
	\begin{align}\label{eq:Guu_restricted}
		G_{uu}^{(2)}=&\frac12(\mathcal R-2)+3k^2+2\partial_u K-\partial_u(D_AU^A)\nonumber\\
		&+\frac14\partial_uC_{AB}\partial_uC^{AB}
		-2(1+k)\partial_u m,\\
		G_{rr}^{(3)} =& -2(1+k)(2ku-K),\\
		G_{ur}^{(2)}=&\frac12(\mathcal R-2)+3k^2,\\
		G_{uA}^{(1)}=&\frac12\left(\partial_uD^BC_{AB}+2\partial_uU_A+\partial_uD_AK\right),\\
		G_{rA}^{(2)}=&\frac12\left[-2(1+2k)U_A-D_BC_A^B\right.\nonumber\\*
		&\left.-(3+2k)D_AK\right]
		,\\
		\label{eq:GAB_restricted}
		G_{AB}^{(0)}&=k\partial_uC_{AB}-q_{AB}[k(k-2)+\partial_u K],
	\end{align}
	together with 
	\begin{align}\label{eq:riccirestriction_simplified}
		R^{(2)}=R^{(2)}_{\text{FLRW}}+\left[\mathcal{R}-2+2\partial_uK\right] \overset{!}{=}R^{(2)}_{\text{FLRW}} \ .
	\end{align}
	From these equations we find the constraints
	\begin{align}
		K=&\frac{8\pi G}{2(1+k)}T_{rr}^{(3)}+2ku\nonumber\\*
		=&\frac{8\pi G}{2(1+k)}(T_{rr}^{(3)}-T_{rr \ \text{FLRW}}^{(3)})\nonumber\\*
		=&\frac{8\pi G}{2(1+k)}\triangle T_{rr}^{(3)} \ ,  \\
		\partial_u K=&\frac12(2-\mathcal{R})=\frac{8\pi G}{2(1+k)}\partial_u(\triangle T_{rr}^{(3)}) \ , \\
		U_A=&-\frac{8\pi G}{(1+k)}T_{rA}^{(2)}-\frac{q^{BM}}{2(1+k)}(D_MC_{AB})\nonumber\\*
		&-\frac{3+2k}{2(1+k)}D_A K \ ,
	\end{align}
	which indicate that $K$ and $U_A$ do not propagate and are completely determined in terms of the sources and other fields.
	
	Next, we examine in detail \eqref{eq:Guu_restricted} and \eqref{eq:GAB_restricted}. Let us begin by decomposing \eqref{eq:GAB_restricted} into trace and traceless components 
	\begin{align}
		q^{AB}G_{AB}^{(0)}&=-2[k(k-2)+\partial_uK]\label{eq:decloupledGAB_restricted_trace} \ ,  \\
		G_{AB}^{(0)}-\frac12 q_{AB}q^{CD}G_{CD}^{(0)}&= k\partial_uC_{AB}\label{eq:decloupledGAB_restricted_traceless} \ .
	\end{align}
	The former equation does not convey special information but the latter tells us that, for $k\neq0$, the time evolution of $C_{AB}$ is constrained by the sources and it is not anymore a field carrying dynamical degrees of freedom at future null infinity $\mathcal{I}^+$. This is a crucial difference with respect to asymptotically flat spacetimes, where $C_{AB}$ only enters the Bondi mass-loss formula \eqref{eq:Guu_restricted} and is unconstrained.
	Finally, looking in detail at Eq.~\eqref{eq:Guu_restricted}, we observe that $m$ enters the mass loss equation with only one time derivative, which would define its evolution as Cauchy data in terms of energy-momentum components.

	Furthermore, after a lengthy computation, it can be shown that $N_A$ also enters the equations of motion for $G_{uA}^{(2)}$ with only one time derivative and is constrained by the energy-momentum tensor. The subleading coefficient $\mathcal{E}$ in $g_{ur}\simeq \mathcal{O}(r^{-2})$ enters as $q_{AB}\partial_u\mathcal{E}$ in $G_{AB}^{(1)}$ but in $G_{rr}^{(4)}$ it appears linearly without derivative and is fully constrained as can be seen from
	\begin{align}
		G_{rr}^{(4)}=&-\frac14 C_{AB}C^{AB}+2\left[3k^2u^2+2\mathcal{E}+K^2\right.\nonumber\\*
		&\left.+k\left(3u^2+2\mathcal{E}-uK+K^2\right)\right] \ .
	\end{align}

	\paragraph{Short summary}
	We observe that for asymptotically decelerating FLRW spacetimes, the dynamics at future null infinity $\mathcal{I}^+$ is completely constrained. 
    This could have been expected taking into account that, in an expanding universe, there is a Hubble scale from which all the modes stop to be oscillating and simply become frozen. Gravitational waves in the IR limit will, therefore, always be beyond the Hubble scale and do not appear as dynamical from the point of view of $\mathcal{I}^+$. Note, however, that this result depends on the choice of boundary conditions and, in particular, of the fall-off behaviour of the energy-momentum tensor.
	
	Let us close this section with two brief comments.
	In the background $\Theta_A=\Phi=0$, the coefficients $K, \mathcal{E}, U_A$ are fully constrained, while $m, N_A, C_{AB}$ are non-propagating and their evolution equations are determined by the sources. These coefficients represent frozen scalar, vector and tensor modes that stop being dynamical at the Hubble scale due to the appearance of well-known friction terms \cite{Mukhanov:2005sc}.  
	In the most general case with $u$-dependent $\Phi$, $\Psi$, $\Theta_A$ and $q_{AB}$, we point out that these four coefficients and/or their time evolution are also completely constrained in terms of the energy-momentum tensor, such that they are again non-propagating.

	The results derived in this section raise the question whether non-trivial infrared structure can be expected in more realistic cosmological settings where expansion and the Hubble scale are present.

	%======================================
	%====== SUPERTRANSLATION CHARGES  =====
	%======================================
	
	\section{Asymptotic charges for supertranslations}
	\label{Sec:charges}
	
	In this section, we will propose asymptotic charges for supertranslations in the absence of Weyl transformations. In fact, this is the setting we explored in detail in section \ref{subsec:onshellnoWeyl}, for which $\Theta_A=\Phi=0$.
 
    We conjecture the expression of supertranslation charges by introducing a physically motivated ansatz and requiring that the charges reproduce the flat limit and obey an abelian algebra, meaning that they are in a faithful representation of the supertranslation algebra.
    
    We start with a simple ansatz given by the standard supertranslation charges in asymptotically flat space (see \emph{e.g.} \cite{Flanagan:2015pxa}) integrated over the comoving sphere
	\begin{equation}\label{eq:supertranslatcharge1}
		\tilde Q_f=\int_{S^2}\sqrt{q}f(x^A)(a^2 m) \ .
	\end{equation}
	We compute the algebra of charges using the definition for integrable charges in \cite{Barnich:2011mi}
	\begin{align}
		\left\{\tilde Q_{f_1}, \tilde Q_{f_2}\right\} =-\delta_{f_2}\tilde{Q}_{f_1}=-\int_{S^2}\sqrt{q}[f_1\delta_{f_2}(a^2 m)] \ .
	\end{align}
	The required variation reads as
	\begin{align}\label{eq:deltaaf2}
		\delta_{f_2}(a^2m)&=a^2\Bigg\{ f_2\partial_u m-\frac{k(k+2)}{2(1+k)^2}\left(D_AD^A+2\right)f_2 \nonumber\\
		&-\frac{1}{4(1+k)}\left[2(\partial_uU^A)(D_Af_2)\right.\nonumber\\
		&\left.-D_A(\partial_uC^{AB}D_Bf_2-\partial_uKD^Af_2)\right]\Bigg\}
	\end{align}
	and, plugging in the equations of motion, it leads to
	\begin{widetext}
	\begin{align}
		\delta_{f_2}(a^2m)&=a^2\Bigg\{ -\frac{f_2\triangle G_{uu}^{(2)}+D^A(\triangle G_{uA}^{(1)}f_2)-[(\mathcal{R}-2)+2\partial_u K]}{2(1+k)}-\frac{k(k+2)}{2(1+k)^2}\left(D_AD^A+2\right)f_2  \nonumber\\
		&\qquad\quad -\frac{\partial_uC^{AB}\partial_uC_{AB}}{8(1+k)}f_2+\frac{1}{4(1+k)}\left[D_AD_B(f_2\partial_uC^{AB})+D_AD^A(f_2\partial_uK)\right]  \Bigg\}.
	\end{align}
	Let us now recall that we only consider supertranslations, which means $\mathcal{R}=2$. As a consequence, Eq.~\eqref{eq:riccirestriction_simplified} tells us that $\partial_u K=0$. This reduces the previous expression to:
	\begin{align}
		\delta_{f_2}(a^2m) =a^2\Bigg\{&-\frac{f_2\triangle G_{uu}^{(2)}+D^A(\triangle G_{uA}^{(1)}f_2)}{2(1+k)}-\frac{k(k+2)}{2(1+k)^2}\left(D_AD^A+2\right)f_2\nonumber\\*
		&-\frac{\partial_uC^{AB}\partial_uC_{AB}}{8(1+k)}f_2 +\frac{1}{4(1+k)}D_AD_B(f_2\partial_uC^{AB})\Bigg\}.
	\end{align}
    \end{widetext}
	The second term in the first line can be reabsorbed by a redefinition of the charge as:
	\begin{empheq}[box=\fbox]{align}\label{eq:supertranslatcharge2}
		Q_f:=&\tilde{Q}_f-\frac{(k+2)}{2(1+k)}\int_{S^2}\sqrt{q}a^2f(x^A)K\nonumber\\
		=&\int_{S^2}\sqrt{q}a^2f(x^A)\left[m-\frac{(k+2)}{2(1+k)}K\right] \ .
	\end{empheq}
	In this way, we obtain
	\begin{align}
		\left\{Q_{f_1},Q_{f_2}\right\} &=-\delta_{f_2}Q_{f_1}\nonumber\\*
		&=\int_{S^2}a^2\sqrt{g_{S^2}}\frac{1}{(1+k)}\Bigg[\frac18f_1f_2\partial_uC^{AB}\partial_uC_{AB}\nonumber\\*
		&-\frac14f_2\partial_u C^{AB}D_AD_Bf_1 \nonumber \\*
		&\qquad +\frac{f_1f_2\triangle G_{uu}^{(2)}-\triangle G_{uA}^{(1)}f_2D^Af_1}{2}\Bigg] \ .
	\end{align}
	The terms in the first line can be absorbed by a modification of the bracket derived in \cite{Barnich:2011mi} for asymptotically flat spacetimes, as follows:
	\begin{empheq}[box=\fbox]{align}\label{eq:newmodifiedbracket}
		\{Q_{f_1}&,Q_{f_2}\}=-\delta_{f_2}Q_{f_1}\nonumber\\*
		&+ \int_{S^2}\sqrt{q}\frac{a^2}{8(1+k)}\partial_uC^{BC}f_2(-\delta_{f_1}C_{BC}) \ .
	\end{empheq}
	The remaining terms are fluxes and non-integrable terms which can either be added to the definition of the charge, making it non-integrable, or cured by redefinition of the bracket.
	In the case in which $\triangle G_{uu}^{(2)}=\triangle G_{uA}^{(1)}=0$, we have a well-defined charge given by Eq.~\eqref{eq:supertranslatcharge2} and the charge bracket in Eq.~\eqref{eq:newmodifiedbracket}. The algebra is abelian and the charges are non-integrable only when $\partial_uC_{AB}\neq0$.
	
	To study the non-conservation of the charges, we use the evolution equation  \cite{Barnich:2011mi} 
	\begin{equation}
		\frac{d}{du}Q_f=\frac{\partial}{\partial u}Q_f+\delta_{1}Q_f.
	\end{equation}
	Contrary to the analysis in flat spacetimes, $\partial Q_f/\partial u$ includes a contribution coming from the $u$-dependent scale factor. As a result, for the setting with $\triangle G_{uu}^{(2)}=\triangle G_{uA}^{(1)}=0$, we obtain
	\begin{widetext}
	\begin{equation}
		\boxed{\frac{d}{du}Q_f=2\frac{H}{a}Q_{f}-\frac{1}{(1+k)}\int_{S^2}\sqrt{q}a^2\left(\frac18f\partial_uC^{AB}\partial_uC_{AB}-\frac14\partial_u C^{AB}D_AD_Bf\right)} \ ,
	\end{equation}
	\end{widetext}
	where $H=\partial_u a$ denotes the Hubble parameter. %Altogether, $\mathcal{H}=H/a$ corresponds to the so-called conformal Hubble parameter.
	
	The first term is new with respect to flat spacetimes and can be interpreted as a Hubble flow of the evolution of the charge. 
	For the concrete case of $f=1$, the first term is positive and the second is negative. As a consequence, the charge $Q_{f=1}$ is not guaranteed to be monotonically decreasing in time. 
    In fact, the term $(H/a)Q_{f=1}$ couples the expansion rate of the FLRW universe with the charge $Q_{f=1}$ and contributes to the time evolution counterbalancing the loss of energy from the gravitational waves. In other words, the quantity $Q_{f=1}$ cannot be strictly interpreted as the FLRW equivalent of the Bondi mass.

	Let us finish this section with some relevant comments:
	\begin{itemize}
		\item Using the charge \eqref{eq:supertranslatcharge2} and the bracket \eqref{eq:newmodifiedbracket}, we have obtained $\{Q_{f_1},Q_{f_2}\}=Q_{[f_1,f_2]=0}=0$ for a subset of metrics compatible with supertranslations, in which $\Phi=\Theta_A=\partial_uK=\triangle G_{uu}^{(2)}=\triangle G_{uA}^{(1)}=0$.
		\item It is of utmost importance to emphasize that, contrary to asymptotically flat spacetimes, $\partial_u C_{AB}$ can be expressed in terms of the energy-momentum tensor components $T_{AB}^{(0)}$ following Eq.~\eqref{eq:decloupledGAB_restricted_traceless}. This means that the notion of Bondi news associated with propagating degrees of freedom is absent. Instead, a matter flux through the boundary takes the place of the Bondi news. When it is vanishing, it renders the charges integrable.
		\item In general, due to the fact that the evolution of all the metric coefficients is determined by the energy-momentum tensor components, we point out that the interpretation of these charges might be very different from that in asymptotically flat spacetimes.
		\item Although the charges we presented are well motivated, we remark that it should be possible to derive them from first principles, \emph{e.g.} using the Barnich-Brandt method \cite{Barnich:2001jy} upon linearizing over the FLRW background.  
		We leave this for future studies.
	\end{itemize}

	%===========================
	%====== DISCUSSION  ========
	%===========================
	\section{Discussion and conclusions}
	\label{Sec:Discussion}
	
	In this paper, we further delved into asymptotically decelerating spatially flat FLRW spacetimes at future null infinity $\mathcal{I}^+$, originally initiated in \cite{Bonga:2020fhx,Rojo:2020zlz}, refined in \cite{Enriquez-Rojo:2021blc} and briefly reviewed in section \ref{Sec:ReviewFLRW}. Herein, we extended the latter by allowing for asymptotic local Weyl diffeomorphisms, which do not preserve the determinant of the metric on the sphere, and we went a step further by studying for the first time the dynamics of these cosmological spacetimes in General Relativity.
	
	Let us summarize the main results of our analysis:
	
	\begin{itemize}
		\item After relaxing the strong Bondi gauge or, equivalently, enabling the change of the determinant of the metric on the sphere, we have shown that asymptotically decelerating spatially flat FLRW spacetimes at future null infinity $\mathcal{I}^+$ admit an asymptotic algebra isomorphic to the Weyl-BMS algebra $\mathfrak{bmsw}$ in asymptotically flat spacetimes \cite{Freidel:2021fxf}. This result differs from the case considered in \cite{Enriquez-Rojo:2021blc,Bonga:2020fhx}, where $\mathfrak{b}_s$, $\mathfrak{bms}_s$ and $\mathfrak{gbms}_s$ are one-parameter deformations of their asymptotically flat counterparts and unveil a cosmological holographic flow at the level of asymptotic algebras. We, thus, find that this flow is trivial if we allow for local Weyl diffeomorphisms, pointing to the fact that $\mathfrak{bmsw}$ is more rigid to deformations than the other extensions of the BMS algebra.
		\item We performed an on-shell analysis of asymptotically decelerating spatially flat FLRW spacetimes at future null infinity $\mathcal{I}^+$ by computing and analyzing the asymptotic Einstein equations. The general pattern and constraints on the metric coefficients are clear. Nonetheless, for the sake of technical simplicity, we explicitly solved the equations for a subclass of metrics compatible with the supertranslation-like sector. Strikingly, we observed that the boundary dynamics is completely constrained by the sources, such that not even the tensor degrees of freedom propagate in contrast to asymptotically flat spacetimes. From a cosmological perspective, this result is consistent with the presence of a Hubble scale in the expanding universes beyond which all dynamics is frozen.
		\item Making use of the on-shell treatment, we obtained well-defined candidates for supertranslation-like charges in some concrete settings. Interestingly, their evolution equation involves a new Hubble term. 
	\end{itemize}
	
	Finally, we comment on open questions and point out future research directions:
	
	\begin{itemize}
		\item When we started this project, we expected to benefit from the richer structure of FLRW spacetimes and, therefore, to explore not only tensor modes (as in asymptotically flat spacetimes) but also scalar and vector modes and their corresponding memories. Nevertheless, our investigation of the Einstein equations revealed the opposite conclusion: all the modes at future null infinity $\mathcal{I}^+$ are constrained by the sources. There are, nonetheless, two caveats worthwhile to be explored. First, we have used General Relativity as gravity theory, while alternative gravity theories might permit richer dynamics for these cosmological spacetimes at $\mathcal{I}^+$. Second, as pointed out in \cite{Enriquez-Rojo:2021blc}, we should have allowed for logarithmic terms in $r$ in the metrics \eqref{eq:metric_allButPhi_uindependent} to include more realistic solutions, such as cosmological black holes. The reason for not including such terms is purely technical, based on the high difficulty of performing their on-shell analysis. However, it might be that including those terms would lead to less restrictive equations of motion.
		\item A very intuitive guideline to follow is extending our machinery to other types of FLRW universes, with a special emphasis on accelerating spatially flat ones, and comparing to the results obtained in this paper. 
		\item It would be very interesting to explore if the Weyl-BMS algebra $\mathfrak{bmsw}$ belongs to a wider multi-parametric family of deformations. The BMS algebra $\mathfrak{bms}$ and the corresponding deformation $\mathfrak{bms}_s$ are members of the family $W(a,b;\bar{a},\bar{b})$ \cite{Safari:2019zmc,Safari:2020pje,Enriquez-Rojo:2021blc}, and the generalized BMS algebra $\mathfrak{gbms}$ and its deformation $\mathfrak{gbms}_s$ lie within the three-parametric family $gW(a,b,\bar{a})$ \cite{Enriquez-Rojo:2021rtv}. It would be very appealing to obtain such a family for $\mathfrak{bmsw}$ and explore its representatives. For a discussion on the physical relevance of exploring families of deformations that interpolate between symmetry algebras obtained from various boundary conditions at various loci (\emph{e.g.} near horizon or asymptotic) and concrete three-dimensional examples, we refer to \cite{Enriquez-Rojo:2021hna}.
		\item We followed an intuitive procedure to obtain supertranslation-like charges. Nevertheless, we expect that it should be possible to derive them explicitly from the Barnich-Brandt method \cite{Barnich:2001jy} by linearizing over an FLRW background. This technical step is worth pursuing in future studies.
		\item Besides, it would be desirable to obtain charges for the superrotation-like and local Weyl sectors. It is a challenging task, even for the global Killing vectors in $S^2$, because it would involve the next order in the $1/r$ expansion of the Einstein equations, which determines the evolution of the angular momentum aspect. We expect that a refinement of the techniques of holographic renormalization developed for asymptotically flat spacetimes \cite{Compere:2018ylh, Freidel:2021fxf} will be very useful in such an endeavour. 
	\end{itemize}

	%===========================
	%====== ACKNOWLEDGEMENTS  ========
	%===========================
	
	\vspace{1em}
	\section*{Acknowledgements}
	The authors thank Geoffrey~Comp\`ere, Ivo~Sachs, Hamid~Safari and Simone~Speziale for very helpful feedback on the manuscript and I.~Kharag for proofreading this paper. 
	The work of M.E.R and T.H. was funded by the Excellence Cluster Origins of the DFG under Germany’s Excellence Strategy EXC-2094 390783311. 
	The research of R.O. is funded by the European Structural and Investment Funds (ESIF) and the Czech Ministry of Education, Youth and Sports (MSMT), Project CoGraDS - CZ.02.1.01/0.0/0.0/15003/0000437.
	R.O. thank the Institut d'Astrophysique de Paris and the Niels Bohr Institute for the hospitality at different stages of this work.

	%===========================
	%====== APPENDICES  ========
	%===========================
	\appendix
	\begin{widetext}
	\section{Asymptotic Lie derivatives}
	\label{Sec:AppLieDer}
	
	In this appendix, we calculate the Lie derivatives of the off-shell metric \eqref{eq:metric_allButPhi_uindependent} with respect to the asymptotic diffeomorphisms \eqref{eq:Srotacc}. These are given by
	\begin{align}
		a^{-2}\Lie{\xi}{g_{uu}}&=2r\left(\Theta^A\partial_u V_A-\partial_u\xi^{r(V)}\right) \nonumber\\
		&+\left[V^AD_A\Phi+\xi^u\partial_u\Phi+2U_A\partial_uV^A-2\partial_u\xi^{r(0)}-2k(1-\Phi)\xi^{r(V)}\right. \nonumber\\
		&\quad\left.+2K\partial_u\xi^{r(V)}-2(1-\Phi)\partial_u\xi^u+2\Theta_A\partial_u\xi^{A(1)}\right] \nonumber \\
		&+\frac{2}{r}\left[\xi^u\partial_u m-k(1-\Phi)\xi^u-\left((1-2k)m-ku(1-\Phi)\right)\xi^{r(V)}\right. \nonumber\\
		&\qquad-k(1-\Phi)\xi^{r(0)}+V^AD_Am+\frac12\xi^{A(1)}D_A\Phi+K\partial_u\xi^{r(0)}-\partial_u\xi^{r(1)}\nonumber \\
		&\qquad\left. + m\partial_u\xi^u+U_A\partial_u\xi^{A(1)}+\Theta_A\partial_u\xi^{A(2)}+N_A\partial_uV^A\right]+\Op(r^{-2}), \label{eq:uuDiff}\\
		a^{-2}\Lie{\xi}{g_{ur}}&=-\left[(1+2k)\xi^{r(V)}+\partial_u\xi^u\right] \nonumber \\
		&+\frac{1}{r}\left[\xi^u\partial_u K+V^AD_AK+K\partial_u\xi^u-\Theta_A\xi^{A(1)}\right. \nonumber \\
		&\qquad\left.+2k\left(u\xi^{r(V)}-\xi^u-\xi^{r(0)}\right)+2kK\xi^{r(V)}\right]+\Op(r^{-2}),\label{eq:urDiff} \\
		a^{-2}\Lie{\xi}{g_{rA}}&=-q_{AB}\xi^{B(1)}-D_A\xi^u+\frac{1}{r}\left(KD_A\xi^u-C_{AB}\xi^{B(1)}-2q_{AB}\xi^{B(2)}\right)+\Op(r^{-2}),\label{eq:rADiff}\\
		a^{-2}\Lie{\xi}{g_{uA}}&=q_{AB}\partial_uV^B r^2\nonumber\\
		&+r\left[(1+2k)\Theta_A\xi^{r(V)}+\Lie{V}{\Theta_A}
		-\partial_A\xi^{r(V)}+C_{AB}\partial_uV^B\right. \nonumber\\
		&\left. \quad+\xi^u\partial_u\Theta_A+\Theta_A\partial_u\xi^u+q_{AB}\partial_u\xi^{B(1)}\right] \nonumber \\
		&+\Big[(2k\Theta_A+\partial_u U_A)\xi^u+(1+2k)\Theta_A\xi^{r(0)}+2k\xi^{r(V)}(U_A-u\Theta_A) \nonumber\\
		&\quad+\Lie{V}{U_A}+\Lie{\xi^{C(1)}}{\Theta_A}-D_A\xi^{r(0)}+KD_A\xi^{r(V)}-(1-\Phi)D_A\xi^u\nonumber \\
		&\quad+\left(\mathcal{D}_{AB}+\frac12C_{AC} C^C_B\right)\partial_uV^B+U_A\partial_u\xi^u+C_{AB}\partial_u\xi^{B(1)}+q_{AB}\partial_u\xi^{B(2)}\Big]\nonumber \\
		&+\frac{1}{r}\Big[\xi^u\partial_u N_A+N_A\partial_u\xi^u+\Lie{V}{N_A}-(1-2k)N_A\xi^{r(V)} \nonumber \\
		&\qquad+KD_A\xi^{r(0)}-D_A\xi^{r(1)}+2mD_A\xi^u+2kU_A(\xi^{r(0)}+\xi^u-u\xi^{r(V)}) \nonumber \\
		&\qquad+2k\Theta_A\left(u^2\xi^{r(V)}-u(\xi^{r(0)}+\xi^u)+\xi^{r(1)}\right)+\Theta_A\xi^{r(1)}+C_{AB}\partial_u\xi^{B(2)} \nonumber\\
		&\qquad+\left(\mathcal{D}_{AB}+\frac12C_{AC} C^C_B\right)\partial_u\xi^{B(1)}+\Lie{\xi^{B(1)}}{U_A}+\Lie{\xi^{B(2)}}{\Theta_A}\Big]+\Op(r^{-2}),\label{eq:uADiff}\\
		a^{-2}\Lie{\xi}{g_{AB}}&=r^2F_{AB}+rS_{AB}+K_{AB}, \label{eq:LieDiffAB2}
	\end{align}
	with 
	\begin{align}
		F_{AB}&=2(1+k)\xi^{r(V)}q_{AB}+\xi^u\partial_u q_{AB}+\Lie{V}{q_{AB}}\nonumber \ , \\
		S_{AB}&=2q_{AB}((1+k)\xi^{r(0)}-ku\xi^{r(V)}+k\xi^u)+\Lie{\xi^{A(1)}}{q_{AB}}\nonumber\\
		&+\Theta_AD_B\xi^u+\Theta_B D_A\xi^u+(1+2k)C_{AB}\xi^{r(V)}+\Lie{V}{C_{AB}}+\xi^u\partial_u C_{AB}\nonumber \ , \\
		K_{AB}&=2kq_{AB}\left(u^2\xi^{r(V)}-u\xi^{r(0)}-u\xi^u\right)+2(1+k)q_{AB}\xi^{r(1)}+\Lie{\xi^{A(2)}}{q_{AB}}\nonumber\\
		&+U_AD_B\xi^u+U_BD_A\xi^u+\Lie{\xi^{A(1)}}{C_{AB}}+2k\left(\mathcal{D}_{AB}+\frac12C_{AC} C^C_B\right)\xi^{r(V)}\cr
		&+\xi^u\partial_u \left(\mathcal{D}_{AB}+\frac12C_{AC} C^C_B\right)+\Lie{V}{\left(\mathcal{D}_{AB}+\frac12C_{AC} C^C_B\right)}\label{eq:LieAB11srot} \ .
	\end{align}
	
	\section{Weyl scalars}
	\label{Sec:Weylscalars}
	
	The Bondi gauge suggests a frame where one can compute the Weyl scalars. This computation has been useful to identify covariant quantities in asymptotically flat spacetimes \cite{Freidel:2021fxf} and we expect that it will also be useful for asymptotically FLRW. For completion, we compute in this appendix the Weyl scalars associated with the on-shell metric \eqref{eq:metric_allButPhi_uindependent}. 
	
	Our starting point is the historical Bondi-Sachs form of the metric
	\begin{align}
		\label{eq:metric_bondiform}
		\dd s^2=-2\eul^{2\beta}a^2\dd u(\dd r+F\dd u)+g_{AB}(\dd x^A-\tilde U^A\dd u)(\dd x^B-\tilde U^B\dd u) \ .
	\end{align}
	The null tetrads are defined by $\eta_{ab}e^a\otimes e^b=g_{\mu\nu}\dd x^\mu\otimes\dd x^\nu$ with
	\begin{align}
		\eta_{ab}&=\begin{pmatrix}
			0 & -1 & 0 & 0\\
			-1 & 0 & 0 & 0\\
			0 & 0 & 0 & 1\\
			0 & 0 & 1 & 0
		\end{pmatrix} \ .
	\end{align}
	For the metric in Eq.~\eqref{eq:metric_bondiform} they are given by
	\begin{align}
		e^0=a\eul^{2\beta}\dd u \ , \quad e^1=a\left(\dd r+F\dd u\right) \ , \quad e^i= a r E^i_A\left(\dd x^A-U^A\right) \ ,
	\end{align}
	with $\eta_{ij}E^i_A E^j_B=\frac{1}{a^2r^2}g_{AB}$ for $i,j\in\{2,3\}$. The corresponding vectors are given by
	\begin{align}
		\hat e_0=\frac{1}{a}\eul^{-2\beta}\left(\partial_u-F\partial_r+U^A\partial_A\right) \ , \quad \hat e_1=\frac{1}{a}\partial_r \ ,
		\quad \hat e_i=\frac{1}{a}\frac{1}{r}\hat E_i^A\partial_A \ .
	\end{align}
	It can be checked easily that the vectors $\hat e_a$ are null. 
	To obtain the metric in the previous form \eqref{eq:metric_allButPhi_uindependent}, we have to expand the parameters in \eqref{eq:metric_bondiform} in the following way:
	\begin{align}
		&\beta=-\frac{K}{2r}-\frac{\mathcal{E}+\frac12K^2}{2r^2}+\Op(r^{-3}) \ , \cr 
		&F=F_0+\frac{F_1}{r}+\frac{F_2}{r^2}+\Op(r^{-3}) \ , \cr
		&F_0=\frac12(1-\Phi+\Theta_A\Theta^A) \ , \cr
		&F_1=(K(1-\Phi)-2m+\Theta^A(K\Theta_A-C_{AB}\Theta^B+2U_A)) \ ,\cr
		&F_2=\frac12\left(\mathcal{E}(1-\Phi)-\mathcal{F}+K(K(1-\Phi)-2m)+2N^A\Theta_A+(\mathcal{E}+K^2)\Theta_A\Theta^A\right.\cr&+
		\left.\frac12(C_A^MC_{BM}-2\mathcal{D}_{AB})\Theta^A\Theta^B+U_A(2K\Theta^A+U^A)-C_{AB}\Theta^A(K\Theta^B+2U^B)\right) \ , \cr
		&\frac{g_{AB}}{a^2}=r^2\left[q_{AB}+\frac{1}{r}C_{AB}+\frac{1}{r^2}\left(\mathcal{D}_{AB}+\frac12C_{AC}C^C_B\right)+\frac{1}{r^3}E_{AB}++\Op(r^{-4})\right] \ , \cr
		&\frac{g_{AB}\tilde U^B}{a^2}=r\Theta_A+U_A+\frac{1}{r}N_A+\Op(r^{-2}) \ ,
	\end{align}
	with $C^A_A=\mathcal D^A_A=0$ to satisfy the determinant condition of the Bondi gauge.\\
	The tetrads on the sphere are expanded as
	\begin{align}
		E^i_A&=\bar E^i_A+\frac{1}{2r}\bar E^i_B C^B_A+\frac{1}{2r^2}\bar E^i_B\left(\mathcal D^B_A+\frac14C_{AC}C^{BC}\right)\cr
		&+\frac{1}{2r^3}\hat{\bar E}^i_B\left(E_A^B-\frac14\left(\mathcal{D}_A^C C_C^B+C_{AC}\mathcal{D}^{CB}\right)-\frac18C_{AC}C^B_D C^{CD}\right)+\Op(r^{-4}) \ , \\
		\hat E_i^A&=\hat{\bar E}_i^A-\frac{1}{2r}\hat{\bar E}_i^BC_B^A-\frac{1}{2r^2}\hat{\bar E}_i^B\left(\mathcal D_B^A-\frac14C^{AC}C_{BC}\right)\cr
		&+\frac{1}{2r^3}\hat{\bar E}_i^B\left(-E_B^A+\frac18C_{B}^DC_{CD}C^{AC}+\frac34\Big(\mathcal{D}^A_C C^C_B+\mathcal{D}_{BC} C^{CA}\Big)\right)
		+\Op(r^{-4}) \ ,
	\end{align}
	where $\hat{\bar E}^A_i$ are the tetrads of the leading term of the metric on the sphere, defined as $q^{AB}=\hat{\bar E}^A_{i}\hat{\bar E}^B_{j}\eta^{ij}$ and $\epsilon^{AB}=\hat{\bar E}^A_{i}\hat{\bar E}^B_{j}\epsilon^{ij}$.
	
	With these tetrads, the Weyl scalars are given by
	\begin{align}
		\Psi_4&=C_{\mu\nu\gamma\delta}\hat{e}^\mu_0\hat{e}^\nu_3\hat{e}^\gamma_0\hat{e}^\delta_3=C_{\hat{0}\hat{3}\hat{0}\hat{3}}
		=a^{-2}\hat{\bar E}^A_3\hat{\bar E}^B_3\left[\frac{1}{r}\psi^{4}_{AB}+\Op(r^{-2})\right],\\
		\Psi_3&=C_{\hat{0}\hat{3}\hat{0}\hat{1}}
		=a^{-2}\hat{\bar E}^A_3\left[\frac{1}{r^2}\Big(\psi^{3}_{A}\Big)+\Op(r^{-3})\right],\\
		\mathrm{Re}(\Psi_2)&=C_{\hat{1}\hat{0}\hat{1}\hat{0}}
		=a^{-2}\left[\frac{1}{r^2}\psi^{2,1}+\frac{1}{r^3}\psi^{2,2}+\Op(r^{-4})\right],\\
		\mathrm{Im}(\Psi_2)&=C_{\hat{1}\hat{0}\hat{2}\hat{3}}
		=a^{-2}\left[\frac{1}{r^2}\tilde\psi^{2,1}+\frac{1}{r^3}\tilde\psi^{2,2}+\Op(r^{-4})\right],\\
		\Psi_1&=C_{\hat{1}\hat{0}\hat{1}\hat{2}}=a^{-2}\hat{\bar E}^A_2\left[\frac{1}{r^3}\psi^{1,1}_A+\frac{1}{r^4}\psi^{1,2}_A+\Op(r^{-5})\right],\\
		\Psi_0&=C_{\hat{1}\hat{2}\hat{1}\hat{2}}=a^{-2}\hat{\bar E}^A_2\hat{\bar E}^B_2\left[
		\frac{1}{r^4}\psi^{0,1}_{AB}+\frac{1}{r^5}\psi^{0,2}_{AB}+\Op(r^{-6})\right],
	\end{align}
	with
	\begin{align}
		\psi^4_{AB}&=-\frac12\partial_u^2C_{AB},\\
		\psi^3_A&=\frac14\left(2D_A\Phi+2\partial_uU_A-\partial_uD_BC^B_A+\Theta_A\mathcal{R}+2\Theta^B(D_B\Theta_A-D_A\Theta_B)\right.\cr
		&\left.+\partial_uD_A K+\Theta_A\partial_u K-\Theta_AD_B\Theta^B+D_BD_A\Theta^B-D_B D^B\Theta_A\right),\\
		\psi^{2,1}&=-\frac16\left(2\Phi+\mathcal{R}-2+D_A\Theta^A+2\partial_u K\right),\\
		\psi^{2,2}&=-2m-\frac16C^{AB}\partial_u C_{AB}-\frac23\partial_u\mathcal{E}-\frac16D_A(2U^A+D_BC^{AB})+\frac13\Theta^A\left(2U_A+D_BC_A^B\right)\cr
		&+K\left(1-\Phi+\frac13\Theta_A\Theta^A-\frac16D_A\Theta^A-\partial_u K\right)-\frac12\Theta^A D_A K+\frac16\Delta K,\\
		\tilde\psi^{2,1}&=\frac14\epsilon^{AB}D_A\Theta_B,\\\
		\tilde\psi^{2,2}&=\frac12\epsilon^{AB}\left(D_AU_B-\frac14C_A^C\partial_u C_{CB}-\frac12C_{AC}D^C\Theta_B+\frac12C_{AC}D_B\Theta^C\right.\cr
		&+\left.\frac12K D_A\Theta_B-\frac12\Theta_AD_B K-\frac12\Theta^CD_AC_{BC}\right),\\
		\psi^{1,1}_{A}&=\frac14\left(2U_A+D_BC^B_A+K\Theta_A-D_A K\right),\\
		\psi^{1,2}_{A}&=\frac12\left(3N_A-\frac14C_A^B(2U_B+D_C C^C_B)+D_B\mathcal{D}^B_A-\frac14C^{BC}D_AC_{BC}\right.\cr
		&\left.-\Theta^B\Big(\mathcal{D}_{AB}+\frac12C_{BC}C^C_A\Big)+\frac34C_{AB}(D^B K-K\Theta^B)
		+\Theta_A(K^2+\mathcal{E})\right.\cr
		&+\left.2K\Big(U_A-\frac12D_A K\Big)-\frac12 D_A\mathcal{E}\right),\\
		\psi^{0,1}_{AB}&=-\mathcal{D}_{AB}-\frac14C_A^C C_{CB}-\frac12C_{AB}K,\\
		\psi^{0,2}_{AB}&=-3E_{AB}+\frac12C_A^C C_B^D C_{CD}+2\mathcal{D}_{CA}C^C_{B}-C_{AB}\mathcal{E}-\mathcal{D}_{AB} K-\frac12C_{AB}K^2 \ .
	\end{align}
	Therefore, we observe that the peeling property is not preserved by this metric ansatz, since the terms $\psi^{2,1}$, $\tilde\psi^{2,1}$, $\psi^{1,1}_{A}$ and $\psi^{0,1}_{AB}$ spoil it. Remarkably, the components $K$ and $\Theta_A$, which are directly determined in terms of the fluid energy-momentum tensor components \eqref{eq:Grr3} and \eqref{eq:GrA1}, are the causant. However, we consistently recover the peeling property in the flat limit, where these four components vanish.
\end{widetext}

\bibliographystyle{apsrev4-1}
\bibliography{references}

\end{document}